# Índice h de las revistas españolas de Historia según *Google Scholar Metrics* (2007-2011)

Emilio Delgado López-Cózar
*EC3: Evaluación de la Ciencia y de la Comunicación Científica*
*Universidad de Granada*
*edelgado@ugr.es*

Manuel Ramírez Sánchez
*Departamento de Ciencias Históricas*
*Universidad de Las Palmas de Gran Canaria*
*mramirez@dch.ulpgc.es*

**RESUMEN**

La aparición de Google Scholar Metrics en abril de 2012 como nuevo sistema de evaluación bibliométrica de revistas científicas a partir del recuento de las citas bibliográficas que estas han recibido en Google Scholar, abre nuevas posibilidades para la medición del impacto de las revistas de humanidades. El objetivo de este trabajo es comprobar el alcance de este producto, a partir del análisis de la cobertura que muestra de las revistas españolas de Historia. Las búsquedas bibliográficas se efectuaron entre el 5 y 6 de diciembre de 2012. Se han identificado 69 revistas, cifra que representa tan solo un 24% de las revistas españolas de Historia. Los valores del índice h alcanzados por las revistas son minúsculos lo que hace que el ranking carezca de poder discriminatorio. Se propone un cambio en el diseño de Google Scholar Metrics para adaptarse a los patrones de producción y citación de la Historia, en particular, y las Humanidades, en general.

**PALABRAS CLAVE**
Google Scholar / Google Scholar Metrics / Revistas / Citas / indicadores bibliométricos / Índice H / Evaluación / Ranking / Historia / Arqueología / Prehistoria / Historia Moderna / Historia Medieval / Historia Antigua / Historia Moderna / Historia Económica

# H Index of History journals published in Spain according to Google Scholar Metrics (2007-2011)

**ABSTRACT**

Google Scholar Metrics (GSM), which was recently launched in April 2012, features new bibliometric systems for gauging scientific journals by counting the number of citations obtained in Google Scholar. This way, it opens new possibilities for measuring journal impacts in the field of Humanities. The present article intends to evaluate the scope of this tool through analysing GSM's searches, from the 5th through 6th of December 2012, of History journals published in Spain. In sum, 69 journals were identified, accounting for only 24% of the History journals published in Spain. The ranges of H index values for this field are so small that the ranking can no longer be said to show a discriminating potential. In the light of this, we would like to propose a change in the way Google Scholar Metrics is designed so that it could also accommodate production and citation patterns in the particular field of History, and, in a broader scope, in the area of Humanities as well.

---

**Referencia bibliográfica recomendada**
Delgado López-Cózar, E.; Ramírez Sánchez, Manuel. Índice H de las revistas españolas de Historia según Google Scholar Metrics (2007-2011). EC3 Working Papers 10: 7 de febrero de 2013

---

# INTRODUCCIÓN

El 15 de noviembre de 2012 Google lanzó una nueva versión de su sistema de evaluación de revistas científicas Google Scholar Metrics (GSM), que había nacido pocos meses antes (abril de 2012), y que ya ha sido objeto de análisis y valoración (1-3). Consciente de las críticas recibidas por la presentación de los rankings de revistas por lenguas y no por disciplinas, en esta nueva versión Google se ha decantado por ofrecer rankings por áreas temáticas y disciplinas. Bien es verdad, que esta opción solo se ha aplicado a las revistas en inglés, quedando excluidas las revistas de los otros nueve idiomas en los que Google presenta listados (chino, portugués, alemán, español, francés, coreano, japonés, holandés e italiano).

Entre las 100 revistas que muestra el listado español de GSM, no figura ninguna revista de Historia. En la actualidad en España circulan en torno a 300 revistas en este campo de conocimiento (288 según RESH; 280 según IN-RECH). Dado el interés despertado por esta nueva herramienta para la evaluación de revistas, y muy especialmente para las de Humanidades, excluidas sistemáticamente de los bases de datos tradicionales cabría preguntarse acerca de cuántas revistas españolas de Historia figuran en el nuevo producto. Hasta ahora se ha indagado sobre la cobertura de revistas españolas de Ciencias Sociales y Jurídicas (4-5), pero no de Humanidades, en general, y de Historia, en particular.

Pues bien, el principal objetivo de este trabajo es identificar todas aquellas revistas españolas de Historia que poseen un índice h calculado en este producto.

# MATERIAL Y MÉTODOS

- Ámbito temático cubierto: revistas científicas españolas de Historia.

- La fuente de datos usada es GSM que cubre sólo revistas que hayan publicado al menos 100 artículos en el periodo 2007-2011 y que hayan recibido alguna cita.

- Las revistas se ordenan según el índice h y a igualdad del mismo de acuerdo con la mediana del número de citas obtenida por los artículos que contribuyen al índice h.

- Las búsquedas se realizaron entre el 5 y 6 de diciembre de 2012.

## RESULTADOS

Se han localizado 69 revistas de Historia, cifra que representa tan solo un 24% de las revistas españolas de Historia.

*Tabla 1.*
*Índice H de las revistas españolas de Historia según Google Scholar Metrics, (2007-2011)*

|    | REVISTAS | H Index | Mediana H |
|----|----------|---------|-----------|
| 1  | Revista de Historia Económica. Journal of Iberian and Latin American Economic History | 9 | 29 |
| 2  | Trabajos de Prehistoria | 7 | 9 |
| 3  | Complutum | 7 | 8 |
| 4  | Investigaciones de Historia Económica | 6 | 6 |
| 5  | Revista de Historia Industrial | 5 | 9 |
| 6  | Palaeohispánica: Revista sobre lenguas y culturas de la Hispania antigua | 5 | 7 |
| 6  | Revista de Indias | 5 | 7 |
| 7  | Ayer | 4 | 6 |
| 8  | Hispania: Revista española de historia | 4 | 5 |
| 8  | Anuario de Estudios Americanos | 4 | 5 |
| 8  | Mélanges de la Casa de Velázquez | 4 | 5 |
| 9  | Munibe. Antropologia-arkeologia | 4 | 4 |
| 9  | Gerión | 4 | 4 |
| 9  | Iberoamericana. América Latina - España - Portugal | 4 | 4 |
| 10 | Pyrenae | 3 | 10 |
| 11 | SAGVNTVM. Papeles del Laboratorio de Arqueología de Valencia | 3 | 5 |
| 11 | Studia Historica. Historia Moderna | 3 | 5 |
| 11 | Asclepio. Revista de Historia de la Medicina y de la Ciencia | 3 | 5 |
| 11 | Revista Complutense de Historia de América | 3 | 5 |
| 12 | Anales de Historia Contemporánea | 3 | 4 |
| 12 | Cuadernos de historia contemporánea | 3 | 4 |
| 12 | Cypsela: revista de prehistòria i protohistòria | 3 | 4 |
| 12 | Boletín Americanista | 3 | 4 |
| 12 | Historia Constitucional | 3 | 4 |
| 12 | Historia Social | 3 | 4 |
| 13 | Revista de Historiografía | 2 | 5 |
| 14 | Revista Hispánica Moderna | 2 | 4 |
| 14 | Clío: Revista de historia | 2 | 4 |
| 15 | Hispania Sacra | 2 | 3 |
| 15 | Veleia: Revista de prehistoria, historia antigua, arqueología y filología clásicas | 2 | 3 |
| 15 | Empúries. Revista de món clàssic i antiguitat tardana | 2 | 3 |
| 15 | Anuario de Estudios Medievales | 2 | 3 |
| 15 | Trocadero. Revista de Historia Moderna y Contemporánea | 2 | 3 |
| 15 | Anuario de Estudios Atlánticos | 2 | 3 |
| 15 | Anuario de Historia de la Iglesia | 2 | 3 |
| 15 | HMIC. Història Moderna i Contemporània | 2 | 3 |
| 16 | Arte, arqueología e historia | 2 | 2 |
| 16 | Cuadernos de historia moderna | 2 | 2 |
| 16 | En la España Medieval | 2 | 2 |
| 16 | Revista atlántica-mediterránea de prehistoria y arqueología social | 2 | 2 |
| 16 | Mainake | 2 | 2 |
| 16 | Historia y Política. Ideas, Procesos y Movimientos Sociales | 2 | 2 |
| 16 | Afers. Fulls de Recerca i Pensament | 2 | 2 |
| 16 | Llull. Boletín de la Sociedad Española de Historia de las Ciencias | 2 | 2 |
| 16 | Randa | 2 | 2 |
| 16 | Revista de Historia Militar | 2 | 2 |
| 17 | Revista Internacional de los Estudios Vascos | 1 | 14 |
| 18 | Alcores: revista de historia contemporánea | 1 | 3 |
| 19 | Espacio, tiempo y forma. Serie II, Historia antigua | 1 | 2 |
| 19 | Pedralbes: revista d'història moderna | 1 | 2 |
| 19 | Revista de arqueología | 1 | 2 |
| 19 | Arkeoikuska: Investigación arqueológica | 1 | 2 |
| 19 | Revista de Historia Actual | 1 | 2 |
| 19 | Castillos de España | 1 | 2 |
| 19 | Educació i Història. Revista d'Història de l'Educació | 1 | 2 |
| 19 | L'Avenç | 1 | 2 |
| 20 | Arqueología, historia y viajes sobre el mundo medieval | 1 | 1 |
| 20 | Estrat crític: revista d'arqueologia | 1 | 1 |
| 20 | Iacobus. Revista de Estudios Jacobeos y Medievales | 1 | 1 |
| 23 | Cuadernos Republicanos | 1 | 1 |
| 20 | Historia 16 | 1 | 1 |
| 20 | Hidalguía | 1 | 1 |
| 20 | Revista de Historia Naval | 1 | 1 |
| 20 | Revista General de Marina | 1 | 1 |
| 20 | Torre de los Lujanes | 1 | 1 |

A la vista de estos resultados podemos ofrecer las siguientes valoraciones:

- El grado de cobertura de las revistas españolas de esta especialidad es ínfimo (24% del universo de revistas españolas circulantes), encontrándose muy por debajo del hallado para las revistas españolas de Ciencias Sociales y Jurídicas (69,8%) (4). Los estrictos criterios de inclusión adoptados por GSM (revistas con más de 100 artículos publicados en los últimos 5 años y que reciban alguna cita), dejan fuera a un importante número de publicaciones, incapaces de alcanzar estos umbrales mínimos.
- Los valores de los índices H obtenidos son muy bajos: el 64% de las revistas posee un índice no superior a 2. Para ponderar adecuadamente estos datos baste compararlos con los índices H alcanzados por las revistas de Historia publicadas en inglés (Tabla 2). El valor promedio del H de las 20 revistas con más impacto duplica (10) al obtenido por las revistas españolas de Historia. El valor mínimo de la revista peor clasificada en el ranking anglosajón (7) es el mismo que el de la revista española que ocupa el segundo lugar. Estos escasos valores del h inutilizan de hecho el ranking a efectos de detectar diferencias entre revistas y es un efecto directo de la reducida ventana de citación de 5 años empleada por GSM. Como ya hemos señalado en trabajos anteriores, la adopción de un marco temporal tan reducido (5 años) "está bien para ciencia y tecnología y algunas disciplinas de ciencias sociales (Ej: Psicología, Economía) y para publicaciones anglosajonas, pero no para las ciencias humanas, jurídicas y revistas de orientación nacional"(8), como es el caso de la Historia.

*Tabla 2.*
*Índice H de las 20 revistas de Historia publicadas en inglés con mayor impacto según Google Scholar Metrics, (2007-2011)*

| REVISTAS | H Index | Mediana H |
|---|---|---|
| The Journal of Economic History | 23 | 31 |
| The American Historical Review | 18 | 22 |
| The Economic History Review | 17 | 28 |
| Comparative Studies in Society and History | 12 | 19 |
| The Journal of American History | 12 | 18 |
| History and Theory | 10 | 15 |
| Law and History Review | 10 | 14 |
| The history teacher | 10 | 14 |
| Radical History Review | 9 | 22 |
| Past & Present | 9 | 11 |
| Journal of Global History | 8 | 16 |
| Journal of Social History | 8 | 11 |
| Journal of World History | 8 | 10 |
| The History of the Family | 8 | 10 |
| Continuity and Change | 8 | 9 |
| History Workshop Journal | 8 | 9 |
| Journal of Urban History | 8 | 9 |
| International Labor and Working-Class History | 7 | 19 |
| The Historical Journal | 7 | 17 |
| Hispanic American Historical Review | 7 | 14 |

- Un análisis de la distribución de revistas por especialidades en el ranking pone de manifiesto que los primeros lugares están copados por las revistas de Historia económica, el mismo fenómeno que se detecta en las revistas en inglés (Tabla 2), de Prehistoria e Historia Antigua, así como por las revistas de Historia de América y, que no es más que un efecto lógico del mayor tamaño y vertebración de las comunidades científicas dedicadas a estos temas. Las revistas de Historia económica pueden recibir citaciones de toda la amplísima comunidad de investigadores sobre economía y empresa. Lo mismo ocurre con el ámbito americanista, cuya comunidad científica se localiza a ambos lados del Atlántico.

## CONCLUSIONES

La restrictiva política de indización de GSM así como la adopción de un marco temporal en nada apropiado a las pautas de citación y de obsolescencia científica de las publicaciones en Historia impide que este producto sea una buena fuente para evaluar el impacto científico de las revistas españolas de historia. Si bien es verdad que identifica algunas revistas nucleares, deja fuera a otras muchas. Es por lo que nos reiteramos en la propuesta de confección de una herramienta que obvie estos problemas (8), introduciendo una ventana de citación de 10 años y una reducción del umbral de producción necesario para figurar en el índice.

## FINANCIACIÓN

## BIBLIOGRAFÍA

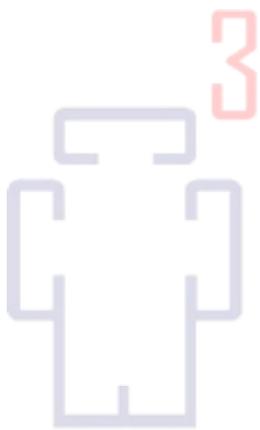